\documentclass{aastex}
\usepackage{emulateapj5,epsfig}

\def\be{\begin{equation}}
\def\ee{\end{equation}}
\def\ba{\begin{eqnarray}}
\def\ea{\end{eqnarray}}
\def\go{\mathrel{\raise.3ex\hbox{$>$}\mkern-14mu
             \lower0.6ex\hbox{$\sim$}}}
\def\lo{\mathrel{\raise.3ex\hbox{$<$}\mkern-14mu
             \lower0.6ex\hbox{$\sim$}}}

\def\beps{{\mbox{\boldmath $\epsilon$}}}
\def\bmu{{\mbox{\boldmath $\mu$}}}
\def\bE{{\bf E}}
\def\bB{{\bf B}}

\def\bI{{\bf I}}
\def\cA{{\cal A}}

\shorttitle{Mode Conversion Due to Vacuum Polarization}
\shortauthors{Lai and Ho}

\begin{document}

\title{Resonant Conversion of Photon Modes Due to Vacuum Polarization
in a Magnetized Plasma: Implications for X-Ray Emission from Magnetars}
\author{Dong Lai and Wynn C.~G. Ho}
\affil{Center for Radiophysics and Space Research,
Department of Astronomy, Cornell University, Ithaca, NY 14853\\
E-mail: dong, wynnho@astro.cornell.edu}

\begin{abstract}
It is known that vacuum polarization can modify the photon 
propagation modes in the atmospheric plasma of a strongly 
magnetized neutron star. A resonance occurs when the effect 
of vacuum polarization on the photon modes balances that of the 
plasma. We show that a photon (with energy $E\go$~a~few keV)
propagating outward in the atmosphere can convert from one polarization mode
into another as it traverses the resonant density, $\rho_{\rm res}\simeq
Y_e^{-1}\eta^{-2}(B/10^{14}~{\rm G})^2(E/1~{\rm keV})^2$~g~cm$^{-3}$, where
$Y_e$ is the electron fraction and
$\eta\sim 1$ is a slowly-varying function of the magnetic field 
$B$. The physics of this 
mode conversion is analogous to the Mikheyev-Smirnov-Wolfenstein 
mechanism for neutrino oscillation. Because the two photon modes 
have vastly different opacities in the atmosphere,
this vacuum-induced mode conversion can significantly affect
radiative transport and surface emission from strongly magnetized 
neutron stars.
\end{abstract}
\keywords{magnetic fields -- radiative transfer -- stars: neutron}

\section{Introduction}

Soft gamma-ray repeaters (SGRs) and anomalous X-ray pulsars (AXPs) are
enigmatic objects. It has been suggested that these are neutron
stars (NSs) endowed  with superstrong magnetic fields, $B\go 10^{14}$~G
(see Thompson \& Duncan 1996). Support for the ``magnetar''
hypothesis has come from several recent observations (e.g., Kouveliotou et
al.~1998; Hulleman et al.~2000; Kaplan et al~2001; Kaspi et al.~2001;
see Hurley 2000 for SGR review; Mereghetti 2000 for AXP review).
Thermal radiation, attributable to surface emission at
$T_{\rm eff}=(3-7)\times 10^6$~K, has already been detected
in four of the five AXPs
and in SGR~1900+14 (see Mereghetti~2000; Perna et al.~2001).
These surface X-ray emissions can potentially reveal much about the 
physical conditions and true nature of SGRs and
AXPs. Theoretical studies on the atmospheres and radiation spectra from
magnetars are still in the early stages (see Ho \& Lai 2001a
and references therein).

In this paper, we are concerned with the effect of vacuum polarization
on the radiative transport in the atmospheric plasma of
strongly magnetized NSs. It is well known that in
strong magnetic fields, vacuum polarization due to virtual $e^+e^-$
pairs can modify the dielectric property of the plasma and the
polarization of photon modes, thereby altering the radiative opacities
(e.g., Adler 1971, Tsai \& Erber 1975; M\'{e}sz\'{a}ros \& Ventura 1979;
Pavlov \& Gnedin 1984; Heyl \& Hernquist 1997; see M\'{e}sz\'{a}ros 1992 for
review). Of particular interest is the ``vacuum resonance'' phenomenon,
which occurs when the vacuum and plasma effects on the linear polarization
of the modes cancel each other, giving rise to a ``resonant'' feature in the
opacities (e.g., Pavlov \& Shibanov 1979; Ventura et al.~1979).
Recent works have begun to address this opacity effect of
vacuum polarization on the radiation spectra from magnetars
(Bezchastnov et al.~1996; Bulik \& Miller 1997; \"Ozel 2001).
At a given density $\rho$ (in g~cm$^{-3}$), the vacuum 
resonance is located at photon energy $E_V
\simeq 1.0\,(Y_e\,\rho)^{1/2}B_{14}^{-1}\eta$~keV, where
$Y_e=Z/A$ is the electron fraction and $\eta\sim 1$ is a
slowly-varying function of the magnetic field $B$ [see eq.~(\ref{eqrhores})].
Because $E_V$ depends on $\rho$ (which spans a wide range in the atmosphere),
one expects that any spectral feature associated with $E_V$ will be
significantly spread out in the spectra of magnetars (Ho \& Lai 2001b).

Here we study a new physical effect overlooked by all previous works:
a photon propagating down the density gradient in the NS
atmosphere can change from one polarization mode to another
at the vacuum resonance. This conversion is particularly effective 
for photons with $E\go$~a few keV. The physics of such mode conversion 
is analogous to the Mikheyev-Smirnov-Wolfenstein (MSW) 
neutrino oscillation which is thought to take place in the Sun (e.g., Bahcall
1989; Haxton 1995)\footnote{
This adiabatic mode conversion is not to be confused with the
physically different (and less drastic) effect of mode
switching due to scattering (see M\'{e}sz\'{a}ros~1992),
which has been included in previous works (e.g., Pavlov et al.~1995;
Ho \& Lai~2001a; \"{O}zel~2001).  \label{foot:mcscat}
}. We show that the mode conversion can dramatically 
affect radiative transport in magnetar atmospheres.

\section{Photon Modes in a Magnetized Electron-Ion
Plasma Including Vacuum Polarization}

In a magnetized plasma + vacuum medium, the dielectric tensor $\beps$
and inverse permeability tensor $\bar\bmu$ can be written as the sums
of the usual plasma terms and the vacuum correction, i.e.,
$\beps=\beps^{(p)}+\Delta\beps^{(v)}$,
$\bar\bmu=\bI+\Delta\bar\bmu^{(v)}$ (where $\bI$ is the unit tensor).
The plasma (electrons and ions) dielectric tensor $\beps^{(p)}$ is given in,
e.g. Shafranov (1967).
The vacuum corrections can be written as
\be
\Delta\beps^{(v)}=(a-1)\bI +q\hat\bB\hat\bB,\qquad
\Delta\bar\bmu^{(v)}=(a-1)\bI+m\hat\bB\hat\bB,
\ee
where $\hat\bB$ is the unit vector along the magnetic field
$\bB$, and $a$, $q$, and $m$ are functions of $B$.
For $B\ll B_Q=m_e^2c^3/e\hbar=4.414\times 10^{13}$~G,
\be
a=1-2\delta,~~q=7\delta,~~m=-4\delta,\quad {\rm with}~~
\delta=\left({\alpha\over 45\pi}\right)b^2,
\ee
where $\alpha=1/137$ and $b\equiv B/B_Q$ (Adler 1971).
For $B\go B_Q$, using the expansion in Heyl \& Hernquist (1997) (see also
Tsai \& Erber 1975), we have
\ba
&&a\simeq 1+{\alpha\over 2\pi}\left[1.195-{2\over 3}\ln b-{1\over b}
(0.855+\ln b)-{1\over 2b^2}\right],\\
&&q\simeq -{\alpha\over 2\pi}\left[-{2\over 3}b+1.272-{1\over b}
(0.307+\ln b)-{0.7\over b^2}\right],\\
&&m\simeq -{\alpha\over 2\pi}\left[{2\over 3}+{1\over b}(0.145-\ln b)-{1\over
b^2}\right].
\ea
An electromagnetic (EM) wave with a given frequency $\omega$ satisfies the
wave equation
\be
\nabla\times(\bar\bmu\cdot\nabla\times\bE)={\omega^2\over c^2}\,\beps\cdot
\bE.
\label{waveeq}\ee
For normal modes propagating along the $z$-axis, with
$\bE\propto e^{ik_\pm z}$, this reduces to
\be
-k_\pm^2\hat z\times\left[\bar\bmu\cdot (\hat z\times \bE_{\pm})
\right]={\omega^2\over c^2}\,\beps\cdot\bE_\pm,
\label{eqmode}\ee
where the subscripts ``$\pm$'' specify the two modes 
(``plus-mode'' and ``minus-mode'').
Using the procedure of M\'{e}sz\'{a}ros \& Ventura (1979; who neglected ions
and considered the $b\ll 1$ limit), the mode eigenvector
$\bE_\pm$ can be expressed explicitly: In the $xyz$ coordinates
with ${\bf k}$ along the $z$-axis and $\bB$ in the x-z plane
(such that ${\hat k}\times\hat\bB=\sin\theta_B\,\hat y$, where
$\theta_B$ is the angle between ${\bf k}$ and $\bB$), we write
$\bE_\pm=\bE_{\pm T}+E_{\pm z}\hat z$, and the transverse part is 
\be
\bE_{\pm T}={1\over (1+K_\pm^2)^{1/2}}(iK_\pm,1),
\ee
where
\be
K_{\pm}=\beta\pm\sqrt{\beta^2+r},
\ee
with $r=1+(m/a)\sin^2\theta_B\simeq 1$, and (assuming $v\ll 1$)
\be
\beta\simeq  {u_e^{1/2}(1-u_i)\sin^2\theta_B\over 2\cos\theta_B}
\left[1-{q+m\over v}\left(1-u_e^{-1}\right)\right].
\label{eqbeta}\ee
Here $u_e=(\omega_{Be}/\omega)^2$, $u_i=(\omega_{Bi}/\omega)^2$,
$v=(\omega_p/\omega)^2$, and the electron cyclotron frequency
$\omega_{Be}$, the ion cyclotron frequency $\omega_{Bi}$, and the electron
plasma frequency $\omega_p$ are given by
$\hbar\omega_{Be} = \hbar eB/(m_ec)=1158\,B_{14}$~keV,
$\hbar\omega_{Bi} =\hbar ZeB/(Am_pc)= 0.6305\,B_{14}$~keV,
$\hbar\omega_p= \hbar (4\pi n_e e^2/m_e)^{1/2}=28.71\,
(Y_e\rho_1)^{1/2}$~eV. We shall be interested in photon energies
such that $u_e\gg 1$ and $v\ll 1$.

The standard way of classifying photon modes in a magnetized plasma
(e.g., M\'{e}sz\'{a}ros 1992) relies on the value of $|K|=|E_x/E_y|$:
the extraordinary mode (X-mode) has $|K|\ll 1$, and its $\bE$ is mostly
perpendicular to the ${\hat k}$-$\hat\bB$ plane;
the ordinary mode (O-mode) has $|K|\gg 1$, and is polarized
along the ${\hat k}$-$\hat\bB$ plane. The advantage of such classification
is that the X-mode and O-mode interact very differently with matter:
the O-mode opacity is largely unaffected by the magnetic field,
while the X-mode opacity is significantly reduced (by a factor
of order $\omega^2/\omega_{Be}^2)$ from the zero-field value.
However, this classification becomes ambiguous when $|\beta|\lo 1$
($|K|\sim 1$). Obviously, $\beta=0$ for $\omega=\omega_{Bi}$. 
But even for general energies ($E\neq \hbar\omega_{Bi}$), a photon 
traveling in an inhomogeneous medium 
encounters $\beta=0$ when the vacuum resonance ($v=q+m$) occurs,
namely at the resonance density
\be
\rho_{\rm res}=0.964\,Y_e^{-1}(B_{14}E_1)^2\eta^{-2}\,{\rm g~cm}^{-3},
\label{eqrhores}\ee
where $E_1=E/(1~{\rm keV})$ and
$\eta=[3\delta/(q+m)]^{1/2}$ is a slow-varying function of
$b$ [$\eta=1$ for $b\ll 1$ and
$\eta\rightarrow (b/5)^{1/2}=0.673\,B_{14}^{1/2}$
for $b\gg 1$; $\eta$ varies from $0.99$ at $B_{14}=1$ to $6.7$ at
$B_{14}=100$].

Figure 1 shows the values of $K_\pm$ and the refractive indices
$n_\pm=ck_\pm/\omega$ for the plus and minus-modes near the resonance
density.  The plus-mode (minus-mode) manifests as O-mode (X-mode)
at high densities but becomes X-mode (O-mode) at low densities.
A natural question arises: how does the polarization state evolve
as a photon traverses from the high density region through $\rho_{\rm res}$
to the low density region? Before we proceed,
it is convenient to introduce the ``mixing'' angle $\theta_m$ via
$\tan\theta_m=1/K_+$, so that
\be
\tan 2\theta_m=\beta^{-1},
\label{eqthetam}\ee
where we have used $r-1\ll 1$. The eigenvectors of the modes are
$\bE_{+T}=(i\cos\theta_m,\sin\theta_m)$ and
$\bE_{-T}=(-i\sin\theta_m,\cos\theta_m)$.
Clearly, at the resonance $\theta_m=45^\circ$, the X-mode and O-mode are
maximally ``mixed''.

\section{Mode Conversion in an Inhomogeneous Plasma}

If the density variation is sufficiently gentle,
an O-mode photon created at high densities will remain
on the $K_+$-trajectory (Fig.~1) as it travels outward
and will adiabatically convert into the X-mode after traversing
the resonance density. This is analogous to the MSW mechanism
for neutrino oscillation. To quantify this, we write the general, 
polarized EM wave with frequency $\omega$ traveling 
in the $z$-direction as
\be
\bE(z)=A_+\bE_+e^{i\phi_+}+A_-\bE_- e^{i\phi_-},
\label{field}\ee
where $\phi_\pm(z)=\int^z\,k_\pm\,dz$.
We shall adopt the WKB (or geometric optics) approximation,
in which the $A_\pm$ and $\bE_\pm$ vary on a length scale
much longer than the photon wavelength. Substituting eq.~(\ref{field})
into eq.~(\ref{waveeq}), using eq.~(\ref{eqmode})
and the approximations $a-1\ll1$ and $m\ll 1$, and then multiplying
the resulting equation by $\bE_{+T}^\ast$ and $\bE_{-T}^\ast$, respectively,
we obtain the amplitude evolution equations
\be
\left(\begin{array}{c}A_+'\\A_-'\end{array}\right)
=\left[\begin{array}{cc}
-{k_+'\over 2k_+} & {k_-\over k_+}\theta_m' e^{-i\Delta\phi} \\
-{k_+\over k_-}\theta_m' e^{i\Delta\phi} & -{k_-'\over 2k_-}
\end{array} \right]
\left(\begin{array}{c}A_+\\A_-\end{array}\right),
\label{eqap}\ee
where $'$ stands for $d/dz$ and $\Delta\phi=\phi_+-\phi_-$.
Neglecting the off-diagonal terms, we find
$A_\pm(z)\propto |k_\pm(z)|^{-1/2}$, which is a standard WKB result.
Defining $\cA_\pm\equiv A_\pm e^{i\phi_\pm}$,
we can cast eq.~(\ref{eqap}) into a form analogous to that of
quantum mechanics:
\be
i\left(\begin{array}{c}\cA_+'\\ \cA_-'\end{array}\right)
\simeq \left(\begin{array}{cc}
-\Delta k/2 & i\theta_m'\\
-i\theta_m' & \Delta k/2
\end{array} \right)
\left(\begin{array}{c}\cA_+\\ \cA_-\end{array}\right),
\label{eqap2}\ee
where $\Delta k=k_+-k_-$, and we have subtracted a nonessential
unity matrix and used $k_+\simeq k_-$ and
$|k_\pm'/k_\pm|\ll |k_\pm|$.
Clearly, when $|\theta_m'|\ll |\Delta k/2|$, or
\be
\gamma\equiv \left|{(n_+-n_-)\omega\over 2\theta_m'c}\right|\gg 1,
\label{eqgamma}\ee
the polarization vector will evolve adiabatically (e.g., a photon
in the plus-mode will remain in the plus-mode). Using
eqs.~(\ref{eqthetam}) and (\ref{eqbeta}), we find
\be
\theta_m'=-{1\over 2}\sin^2 2\theta_m {u_e^{1/2}(1-u_i)\sin^2\theta_B
\over 2\cos\theta_B}\left({\rho_{\rm res}\over\rho}\right){\rho'\over\rho}.
\ee
The difference in refractive indices of the two modes is
\be
n_+-n_-\simeq -{v\cos\theta_B\over u_e^{1/2}(1-u_i)\sin 2\theta_m}.
\ee
Clearly, the adiabaticity condition (\ref{eqgamma}) is most easily
violated at the resonance ($\theta_m=45^\circ$).
Evaluating eq.~(\ref{eqgamma}) at $\rho=\rho_{\rm res}$, we obtain
\be
\gamma_{\rm res}=(E/E_{\rm ad})^3,
\ee
\be
E_{\rm ad}=2.52\,\left(\eta\,\tan\theta_B\right)^{2/3}|1-u_i|^{1/3}
\left({1~{\rm cm}\over H_\rho}\right)^{1/3}~{\rm keV},
\ee
where $H_\rho=|dz/d\ln\rho|$ is the density scale height along the ray
(evaluated at $\rho=\rho_{\rm res}$).
For an ionized hydrogen atmosphere, $H_\rho\simeq
2kT/(m_p g\cos\theta)=1.65\,T_6/(g_{14}\cos\theta)$~cm,
where $T=10^6T_6$~K is the temperature, $g=10^{14}g_{14}$~cm~s$^{-2}$ is
the gravitational acceleration, and $\theta$ is the angle between the ray and
the surface normal. Thus, for $E\go 2E_{\rm ad}$, resonant
conversion between X-mode and O-mode is essentially complete;
for $E\lo 0.5E_{\rm ad}$, a photon will ``jump'' across the adiabatic
curves (Fig.~1), and an X-mode (O-mode) will remain an X-mode (O-mode)
in passing through the resonance. In general, the jump probability
can be calculated with the Landau-Zener formula
\be
P_{\rm jump}=e^{-\pi\gamma_{\rm res}/2}
\ee
(see Haxton 1995). The width of the resonance region
can be estimated by considering $|\beta|=1$ as defining
the edge of the resonance; this gives
\be
(\Delta z)_{\rm res}\simeq 1.7\times 10^{-3}|1-u_i|^{-1}
{\cos\theta_B\over\sin^2\theta_B}\left({E_1\over B_{14}}
\right)H_\rho,
\ee
which is much smaller than the photon mean-free path.

\section{Implications for Surface Emission from Magnetars}

To understand qualitatively the effects of vacuum polarization (and
mode conversion, in particular) on the radiation spectra from magnetar
atmospheres, we begin by estimating the location of the
decoupling layer (at which the optical depth is $2/3$)
for photons of different energies and polarizations.
For simplicity, we neglect scattering and consider fully-ionized
H atmospheres.

First, let us turn off the vacuum polarization and proton effects.
The free-free absorption opacity can be written as $\kappa=\kappa_0\xi$,
where $\kappa_0=9.3\,\rho_1T_6^{-1/2}E_1^{-3}G$~cm$^2$g$^{-1}$ (with
$G=1-e^{-E/kT}$) is the zero-field opacity and $\xi\sim 1$ for the O-mode
and $\xi\sim u_e^{-1}$ for the X-mode (we set the Gaunt factor to unity).
Hydrostatic equilibrium yields the column density $y\simeq 0.83\,\rho_1
T_6$~g~cm$^{-2}$ (for $g_{14}=2$).
The decoupling densities of the two modes are then
$\rho_O^{\rm (nv)}\simeq 0.42\,T_6^{-1/4}E_1^{3/2}G^{-1/2}$~g~cm$^{-3}$
and $\rho_X^{\rm (nv)}\simeq \rho_O\,u_e^{1/2}$
[see Fig.~2; the superscript ``(nv)'' stands for ``no vacuum'';
note that we have treated the temperature as a constant; this a good
approximation when estimating the decoupling density since $T$ varies
by only a factor of a few while $\rho$ varies by many orders of
magnitude above the decoupling layer]. The X-mode photons emerge
from deeper in the atmosphere, and thus they are the main carriers of 
the X-ray flux.

Next we include vacuum polarization and protons, but turn off
mode conversion (e.g., X-mode remains X-mode across the vacuum
resonance). The decoupling density $\rho_O$ for the O-mode
is still $\rho_O^{(\rm nv)}$. For the X-mode, when $\rho_X^{\rm (nv)}\go
\rho_{\rm res}$, or
\be
E\lo E_{c2}\simeq 63\,T_6^{-1/6}(B_{14}/\eta^2)^{-2/3}~{\rm keV},
\ee
the photons created at $\rho_X^{\rm (nv)}$ will encounter the resonance,
near which the X-mode opacity is greatly enhanced; thus the decoupling
density $\rho_X^{\rm (nc)}$ [``(nc)'' stands for ``no conversion'']
will be smaller than $\rho_X^{(\rm nv)}$ (see Fig.~2; 
for $E\go E_{c2}$, $\rho_X^{\rm (nc)}$
is close but not equal to $\rho_X^{\rm (nv)}$
because the opacity is modified slightly by the vacuum effect
even away from the resonance). In fact, the optical depth across
resonance region (say $0.9<\rho/\rho_{\rm res}<1.1$)\footnote{
The X-mode opacity is $\kappa=\kappa_0\xi$, with $\xi
=\sin^2\theta_B/(2\beta)^2+u_e^{-1}$ for $|\beta|\gg 1$ and
$\xi=\sin^2\theta_B/2$ for $|\beta|\ll 1$ (see Ho \& Lai 2001a).
For $|x|\equiv|\Delta\rho/\rho_{\rm res}|\lo 0.1$, 
we have $\beta\sim u_e^{1/2}x$. Thus $\xi\sim 1/(u_ex^2)$ 
for $u_e^{-1/2}\ll |x|\lo 0.1$ and $\xi\sim 1$ for $|x|\ll u_e^{-1/2}$ 
(for typical $\theta_B$'s).  Note that the $|x|\lo 2u_e^{-1/2}$ region 
contributes most to $\Delta\tau$.}
is a factor $\sim u_e^{1/2}$ larger than that of the non-resonant
region ($\rho\lo 0.9 \rho_{\rm res}$). Thus,
as $E$ decreases below $E_{c2}$, we expect $\rho_X^{\rm (nc)}$
to follow closely $\rho_{\rm res}$, until $E$ drops below another critical 
energy $E_{c1}$ (see Fig.~2), which is set by $\Delta\tau\simeq 2/3$, or
\be
E_{c1}\simeq 6\,T_6^{-1/4}B_{14}^{-3/2}\eta^2~{\rm keV}
\ee
(the numerical prefactor $6$ is for $\theta_B=45^\circ$).
Below $E_{c1}$, we find that $\rho_X^{\rm (nc)}$ becomes
close to $\rho_X^{(\rm nv)}$ again, except for the proton cyclotron feature at
$E_{Bi}=0.63\,B_{14}$~keV (studied previously by Ho \& Lai 2001a and Zane et
al.~2001).

Now consider the effect of mode conversion at the vacuum resonance.
For $E>1.3\,E_{\rm ad}$, adiabatic mode conversion is nearly complete
($P_{\rm jump}<3\%$). The O-mode photons traveling from high densities
through $\rho_{\rm res}$ are converted to X-mode photons, which then
freely stream out of the atmosphere. Thus for
$1.3\,E_{\rm ad}\lo E\lo E_{c2}$,
the effective decoupling density for X-mode is $\rho_X=\rho_{\rm res}$.
For $E\go E_{c2}$, vacuum resonance occurs inside both the X and O-mode
decoupling layers, and therefore $\rho_X\simeq \rho_X^{\rm (nv)}$.
For $E\ll E_{\rm ad}$, mode conversion is ineffective, and thus
$\rho_X=\rho_X^{\rm (nc)}$. Around $E_{\rm ad}$, the X-mode photons
are emitted from both $\rho_X^{\rm (nc)}$ (with probability $P_{\rm jump}$)
and $\rho_{\rm res}$ [with probability $(1-P_{\rm jump})$].

To translate Fig.~2 into emergent spectra from NS atmospheres, 
we need to know the temperature profile $T(\rho)$. Obviously, 
$T(\rho)$ can only be determined
from self-consistent atmosphere modeling, and different photon
decoupling behaviors will give rise to different temperature
profiles. This is beyond the scope of this paper. Here,
for illustrative purposes, we consider a fixed temperature profile,
obtained from the $B=10^{14}$~G, $T_{\rm eff}=5\times 10^6$~K,
hydrogen atmosphere model (which includes the proton effect but not vacuum
polarization) of Ho \& Lai (2001a; see Fig.~4 in that paper); this
profile is strongly non-Eddington, showing a
mild plateau around $\rho=2-20$~g~cm$^{-3}$ ($T_6$ varies from 3.5 to 3.7),
as a result of radiative transport by two modes with vastly different
opacities. Given this $T(\rho)$, we calculate the emergent spectral flux by
$F_\nu\simeq (\pi/2) \left\{B_\nu[T(\rho_X)]+B_\nu[T(\rho_O)]\right\}$,
where $B_\nu(T)$ is the Planck function and $\rho_X,\,\rho_O$ are
given as in Fig.~2. In the case of partial mode
conversion, we replace $B_\nu[T(\rho_X)]$ by $(1-P_{\rm jump})
B_\nu[T(\rho_{\rm res})]+P_{\rm jump}B_\nu [T(\rho_X^{\rm (nc)})]$.
Figure 3 shows the results for the different cases discussed
in the previous paragraphs. We see that the vacuum polarization
reduces the flux in the energy band $E_{\rm ad}\lo E\lo E_{c2}$.
Part of the depletion in Fig.~3 is caused by our
assumption of fixed temperature profile (e.g., the heavy-solid
curve and the dashed curve have different total fluxes), but
Fig.~3 is indicative of the potentially important effects
of vacuum polarization (and mode conversion) on the radiation
spectra from magnetars.

For $\rho_O^{\rm (nv)}\go\rho_{\rm res}$, i.e., $B_{14}\lo 0.66\,
Y_e^{1/2}E_1^{-1/4}$, the vacuum resonance occurs outside the O and X-mode
decoupling layers, thus the total spectral flux ($F_X+F_O$) is largely
unaffected by the vacuum resonance, although the polarization will still be
affected.

Previous studies (e.g., \"{O}zel 2001; see footnote~\ref{foot:mcscat})
neglected adiabatic mode
conversion as discussed in this paper.  We expect that including
this effect in the radiative transport would produce a qualitatively
different spectrum for magnetars (see Ho \& Lai 2001b for details).

\acknowledgments
This work is supported in part by NASA grant NAG 5-8484 and
an Alfred P. Sloan fellowship to D.L.


\clearpage

\begin{figure}
\plotone{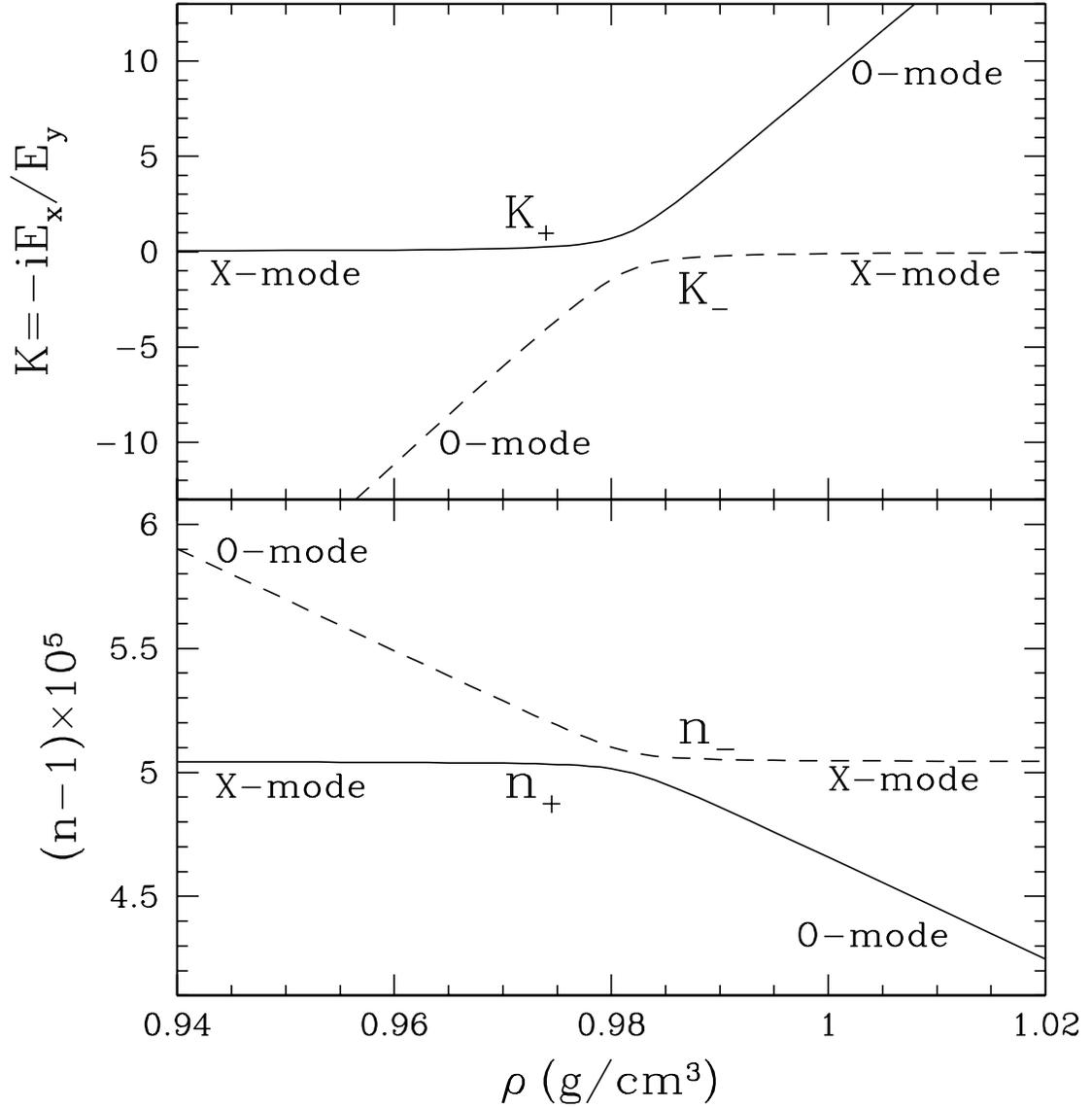}
\caption{The polarization parameters (upper panel) and refractive indices
(lower panel) of photon modes as functions of density near the vacuum
resonance for $B_{14}=1,~E=1$~keV, $Y_e=1$, and $\theta_B=45^\circ$.
\label{fig1}}
\end{figure}

\clearpage
\begin{figure}
\plotone{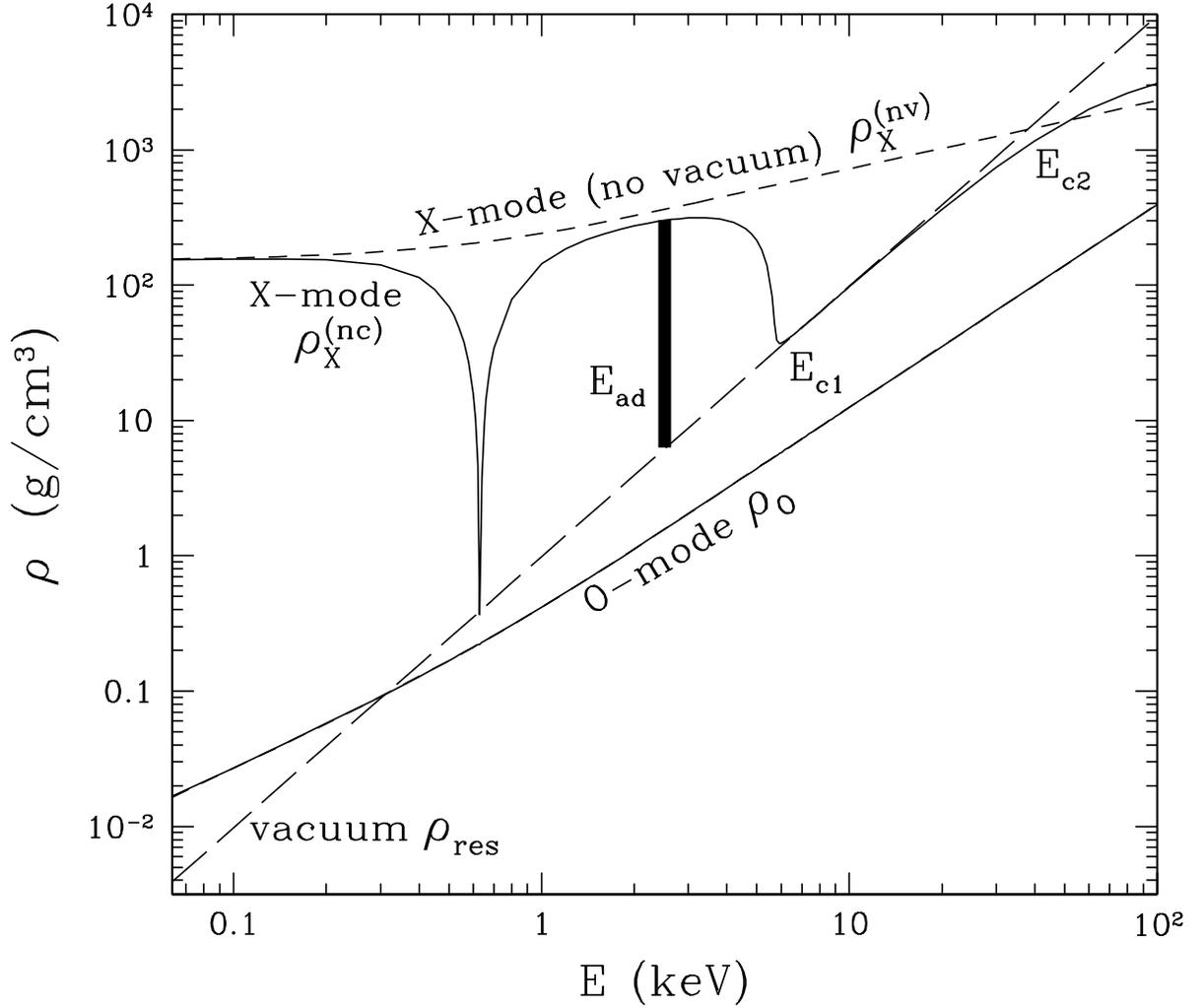}
\caption{The decoupling densities (at which $\tau=2/3$)
of different polarization modes as a function of photon energy for
$B_{14}=1,~\theta_B=45^\circ$, and $T\simeq 5\times 10^6$~K.
The solid lines show the case when vacuum polarization is included
but mode conversion is turned off; the short-dashed line shows the case
when the vacuum polarization and proton effects are
turned off (note $\rho_O$ is unaffected by vacuum effect);
the long-dashed line shows the vacuum resonance density.
The thick vertical line corresponds to the critical energy $E_{\rm ad}$ for
adiabatic mode conversion.
\label{fig2}}
\end{figure}

\clearpage
\begin{figure}
\plotone{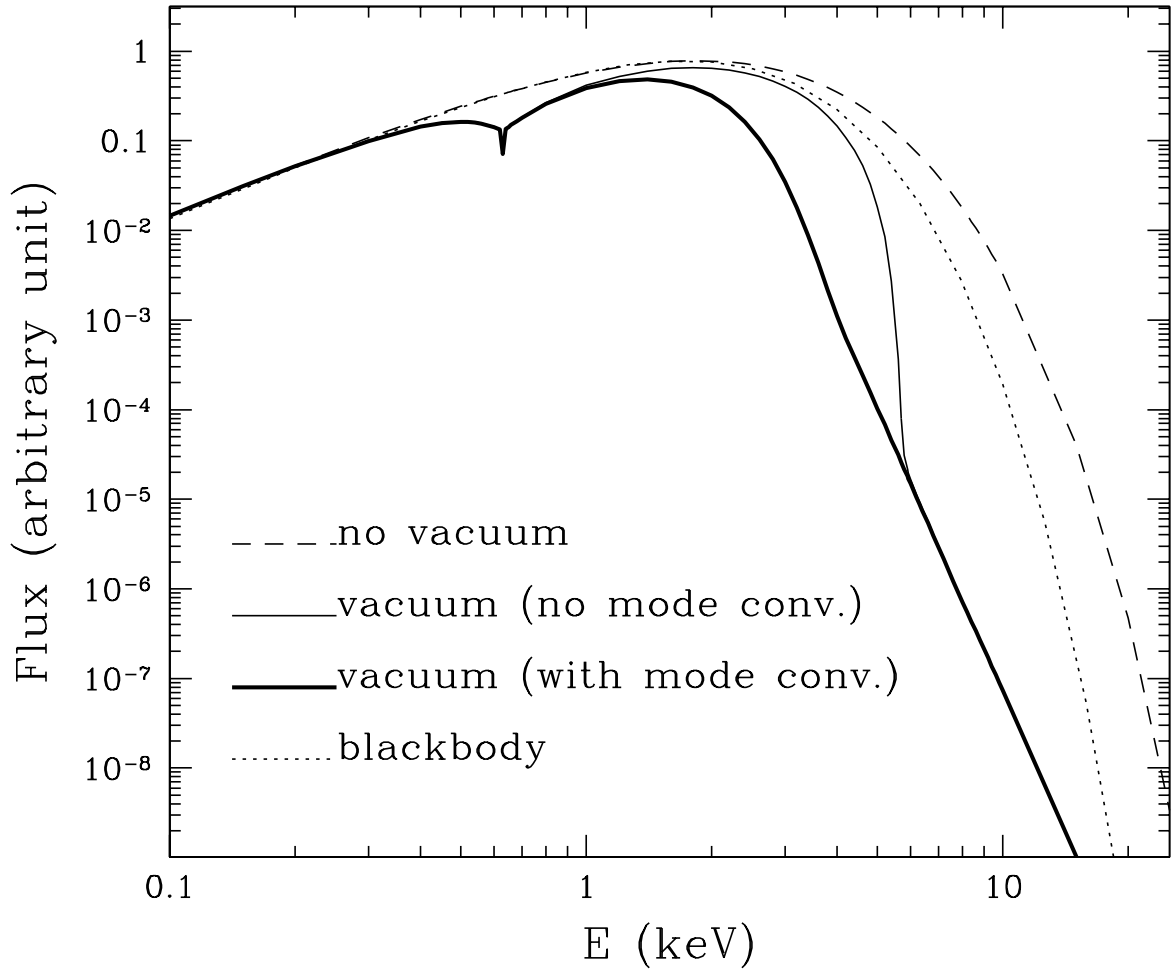}
\caption{Spectral fluxes from NS atmospheres at $B_{14}=1$.
All models assume the same temperature profiles based on  
the $T_{\rm eff}=5\times 10^6$~K model of Ho \& Lai (2001a).
Note that the results are deduced from Fig.~2 (see text) and are
qualitative; they serve to illustrate the effects of 
vacuum polarization and mode conversion. 
\label{fig3}}
\end{figure}

\end{document}